\documentclass[nofootinbib,superscriptaddress,a4paper, twocolumn, floatfix]{revtex4-2}
\usepackage[english]{babel}
\usepackage[utf8]{inputenc}
\usepackage[colorinlistoftodos, color=green!40, prependcaption]{todonotes}
\usepackage{amsthm}
\usepackage{mathtools}
\usepackage{physics}
\usepackage{braket}
\usepackage{listings}
\usepackage{xcolor}
\usepackage{graphicx}
\usepackage[left=23mm,right=13mm,top=35mm,columnsep=15pt]{geometry} 
\usepackage{adjustbox}
\usepackage{placeins}
\usepackage[T1]{fontenc}
\usepackage{lipsum}
\usepackage{csquotes}
\usepackage[pdftex, pdftitle={Article}, pdfauthor={Author}]{hyperref} 
\usepackage{booktabs}
\usepackage{amssymb}
\usepackage{bm}
\usepackage[ruled,vlined]{algorithm2e}
\usepackage[caption=false]{subfig}

\begin{document}

\title{Hamiltonian based graph states ansatz for variational quantum algorithms}
\author{Abhinav Anand}
\email[E-mail:]{abhinav.anand@duke.edu}
\affiliation{Duke Quantum Center, Duke University, Durham, NC 27701, USA.}
\affiliation{Department of Electrical and Computer Engineering, Duke University, Durham, NC 27708, USA.}

\author{Kenneth R. Brown}
\email[E-mail:]{kenneth.r.brown@duke.edu}
\affiliation{Duke Quantum Center, Duke University, Durham, NC 27701, USA.}
\affiliation{Department of Electrical and Computer Engineering, Duke University, Durham, NC 27708, USA.}
\affiliation{Department of Physics, Duke University, Durham, NC 27708, USA.}
\affiliation{Department of Chemistry, Duke University, Durham, NC 27708, USA.}
\date{\today}

\begin{abstract}
    One promising application of near-term quantum devices is to prepare trial wavefunctions using short circuits for solving different problems via variational algorithms.
    For this purpose, we introduce a new circuit design that combines graph-based diagonalization circuits with arbitrary single-qubit rotation gates to get Hamiltonian-based graph states ans\"atze (H-GSA).
    We test the accuracy of the proposed ansatz in estimating ground state energies of various molecules of size up to 12-qubits.
    Additionally, we compare the gate count and parameter number complexity of the proposed ansatz against previously proposed schemes and find an order magnitude reduction in gate count complexity with slight increase in the number of parameters. 
    Our work represents a significant step towards constructing compact quantum circuits with good trainability and convergence properties and applications in solving chemistry and physics problems.
\end{abstract}

\maketitle

\section{Introduction}
As we continue to make progress towards building large quantum computers, we simultaneously need to keep looking for classically intractable problems that can be solved using these devices.
A class of problems~\cite{bharti2022noisy, cerezo2020variational, cao2019quantumreview,mcardle2020quantumreview} that is amenable to currently available noisy device is using them to prepare trial wavefunctions for solving different chemistry and physics problems variationally.
While various proposals~\cite{wecker2015progress,peruzzo2014variational,Anand2021Quantum,kandala2017hardware,tang2021qubit, grimsley2019adaptive, ryabinkin2018qubit, ryabinkin2020iterative} for preparing such states have been put forward in the last few years, there remain many outstanding challenges~\cite{mcclean2018barren,cerezo2021cost,wang2021noise,cerezo2022challenges,anand2022exploring,McClean2016theoryofvqe,anschuetz2022beyond} that need to be addressed.

A common challenge is constructing shallow quantum circuits with good trainability and convergence properties~\cite{mcclean2018barren,cerezo2021cost,wang2021noise}.
Recent advances in understanding of design~\cite{sim2019expressibility,benedetti2019parameterized,cong2019quantum,zhang2022differentiable, du2022quantum} and trainability~\cite{mcclean2018barren,cerezo2021cost,wang2021noise} of different circuits have led to some proposals for constructing such circuits.
One direction is finding better optimization strategies~\cite{skolik2021layerwise, grant2019initialization,anand2021natural,anand2022information} which can overcome some of the shortcoming of the circuit design.
Another direction is to find new design principles to overcome this challenge.
These include using adaptive techniques~\cite{tang2021qubit, grimsley2019adaptive, ryabinkin2018qubit, ryabinkin2020iterative, kottmann2021feasible}, symmetry-breaking ans\"atze~\cite{choquette2020quantum,anand2023leveraging}, and using Clifford or near-Clifford unitaries~\cite{schleich2023partitioning,ravi2022cafqa,anand2022quantum, anand2024stabilizer}, among others~\cite{letcher2023tight}.

In this paper, we focus on the latter and introduce a new class of quantum circuits that we term ``Hamiltonian-based graph states ansatz" (H-GSA).
The proposed ansatz uses graph-based diagonalization circuits to incorporate problem related information and symmetry breaking arbitrary single qubit rotation layer for faster convergence.
This leads to an order of magnitude reduction in the gate count of the proposed ansatz while maintaining the good convergence properties of the Hamiltonian variational ansatz.
We present detailed numerical analysis of our ansatz for estimating ground state energies of different molecules of size up to 12 qubits and find that with a single layer of the ansatz we can get within desired accuracy for many configurations of the molecules considered in this study.
This suggests that our work can be useful in many applications where one requires shallow quantum circuits with good trainability properties, especially when the current devices have limited coherence times.

The remaining sections of this paper are organized as follows: 
We present some preliminary information and the proposed method in sec.~\ref{sec:method}.
The results from different numerical simulations are presented in sec.~\ref{sec:simulations}, and finally, we provide some concluding remarks in sec.~\ref{sec:conclusion}.

\section{Methodology}\label{sec:method}
In this section, we review some background information and present the details of how to construct a Hamiltonian-based graph states ansatz (H-GSA).

\subsection{Hamiltonians}
In the second-quantized formalism, the electronic Hamiltonian is expressed as
\begin{equation}\label{eq:sec_q_ham}
     \hat{H} = \sum_{pq}^{n} h_{pq} \hat{a}^{\dagger}_p \hat{a}_q + \frac{1}{2}\sum_{pqrs}^{n} h_{pqrs} \hat{a}^{\dagger}_p \hat{a}^{\dagger}_q \hat{a}_r \hat{a}_s,
\end{equation}
where, $n$ is the number of spin orbitals, $p,q,r,s$ are spin-orbital indices, $\hat{a}^{\dagger}$ and $\hat{a}$ are creation and annihilation operators, respectively, and the coefficients $h_{pq}$ and $h_{pqrs}$ are one- and two-electron integrals. 
The second-quantized Hamiltonian can be transformed to qubit operators using various mappings, such as Bravyi-Kitaev~\cite{bravyi2002fermionic} or Jordan-Wigner~\cite{jordan1928pauli}.
The resulting Hamiltonian is of the form
\begin{equation}\label{eq:qubit_ham}
    \hat{H} = \sum_{k=1}^{M} c_k \hat{P}_k,
\end{equation}
where $c_k$ is a complex number and $\hat{P}_k$ is a Pauli-string ($\{ \hat{I}, \hat{\sigma}_x, \hat{\sigma}_y, \hat{\sigma}_z \}^{\bigotimes n}$) on $n$-qubits.

\subsection{Groups}
Given the Hamiltonian of the form in Eq.~\ref{eq:qubit_ham}, we now explain how to construct different groups using the Pauli-strings in the Hamiltonian.

\subsubsection{Commuting groups}\label{subsec:CG}
We first review the commutativity property of the terms present in the Hamiltonian. 
Given, two Pauli-strings on $n$-qubits, $\hat{P}_a = \bigotimes^{n-1}_{i=0} \hat{\sigma}^{a}_i$ and $\hat{P}_b  = \bigotimes^{n-1}_{i=0} \hat{\sigma}^{b}_i$, where $\hat{\sigma} \in \{ \hat{I}, \hat{\sigma}_x, \hat{\sigma}_y, \hat{\sigma}_z \}$, we say $\hat{P}_a$ and $\hat{P}_b$ commute if,
\begin{equation}
    [\hat{P}_a, \hat{P}_b] = \hat{P}_a \hat{P}_b - \hat{P}_b \hat{P}_a = 0.
\end{equation}
This can be further divided into two types, qubit-wise commuting (QWC) and general commuting (GC), as follows:
\begin{align}
    [\hat{P}_a, \hat{P}_b] &= \bigotimes^{n-1}_{i=0} \hat{\sigma}_i^a \hat{\sigma}_i^b -  \bigotimes^{n-1}_{i=0} \hat{\sigma}_i^b \hat{\sigma}_i^a \nonumber\\
    &= 0 \begin{cases}
        \text{QWC -} &\text{if $[\hat{\sigma}_{i}^{a}, \hat{\sigma}_{i}^{b}] = 0, \forall i \in \{0,..,n-1\}$, }\\
        \text{GC -} &\text{if the number of indices $i \in \{\mathbb{I}\}$}\\
        &\text{$\subseteq\{0,..,n-1\}$, such that: }\\
        &\text{$[\hat{\sigma}_{i}^{a}, \hat{\sigma}_{i}^{b}] \ne 0$, is even.}
    \end{cases} 
\end{align}
It is also well known~\cite{jena2019pauli} that given two commuting terms, $\hat{P}_{a}$ and $\hat{P}_{b}$, there exist a Clifford circuit, $\mathcal{U}$, that simultaneously diagonalizes them.
The choice of the commutativity type - GC vs QWC - influence the complexity of the Clifford circuit and the number of mutually commuting groups.
 
 In the last few years, several proposals~\cite{verteletskyi2020measurement, izmaylov2019unitary,jena2019pauli, crawford2021efficient, huggins2021efficient,gokhale2020n, zhao2020measurement} have been put forward to find mutually commuting groups within the Hamiltonian. 
 In this article, we cluster the Hamiltonian in $m$ groups of general commuting (GC) terms by first putting all terms in the Hamiltonian of the form, $\hat{P}_j  = \{ \hat{I}, \hat{\sigma}_z \}^{\bigotimes n}$ in a single group and then following the techniques presented in Ref.~\cite{crawford2021efficient} for the rest of the terms.
 Thus, given a Hamiltonian $\Hat{H}$ of the form in Eq.~\ref{eq:qubit_ham}, we can construct $m$ groups, $\mathcal{G}_j \equiv \{ \hat{P}_{j_{k}} \}$, such that
\begin{equation}
    [\hat{P}_{j_{k}} , \hat{P}_{j_{l}}] = 0, \forall \hat{P}_{j_{k}}, \hat{P}_{j_{l}} \in \mathcal{G}_j.
\end{equation}
It is known~\cite{sarkar2019sets} that the maximum number of elements in any of these groups is $2^{n}$, as we ignore the sign of the terms when forming these groups.
It is also known that the maximum number of independent generators of such groups is $n$.
Given such a group $\mathcal{G}_j$, we describe in detail the construction of a corresponding stabilizer group, $\mathcal{S}_j$ in the next section.
 
\subsubsection{Stabilizer groups}\label{subsec:SG}
We first review the properties of a stabilizer group.
A group, $\mathcal{S}$, is a stabilizer group if it is an abelian subgroup of the n-qubit Pauli group, $\mathcal{P}^n$, and $-\hat{I} \notin \mathcal{S}$.
The simultaneous +1 eigenstate of the operators in the group, $\mathcal{S}$, defines the state stabilized by the group.
We use the fact that a product state (the Hartree-Fock state in particular) is an eigenstate of diagonal Pauli-strings to construct stabilizer groups from a set of mutually commuting operators, by following the steps mentioned below.

First, we modify the operators as follows: 
\begin{enumerate}
    \item Given an abelian group, $\mathcal{G}_{j}$, find a Clifford unitary, $\mathcal{U}_j$, which diagonalizes the whole group.
    \item For every operator, $\hat{P}$, in the group, $\mathcal{G}_{j}$, replace it by $-\hat{P}$ if,
    \begin{equation}
        \mathcal{U}_j \hat{P} \mathcal{U}_j^{\dagger} \ket{HF} = - \ket{HF},
    \end{equation}
    where, $\ket{HF}$ is the Hartree-Fock state on $n$-qubits.
\end{enumerate}
After this step, we have another abelian group, $\mathcal{G}_{j}^{'}$, which stabilizes the state $\ket{\Psi_{s}} = \mathcal{U}^{\dagger}_{j} \ket{HF}$.

Next, we convert the operators to their tableau representations, where every operator, $\hat{P}$, is represented as a $2n+1$ dimensional binary vector, $\hat{P} = [X|Z|s]$.
Here, $s \in \{0, 1\}$ represents the sign $(-1)^{s}$ of $\hat{P}$ and $X$ and $Z$ are $n$-dimensional binary vectors, such that $(X_{j}, Z_{j}, P)$ represent the $j$-th component of the operator, $\hat{P}$. 
For instance, $\hat{\sigma}_{x}\hat{I}\hat{\sigma}_{z}\hat{\sigma}_{y} \equiv [1001|0011|0]$.

Once we have the tableau representation of the abelian group, $\mathcal{G}_{j}^{'}$, one can use the Gram-Schmidt procedure~\cite{d2021introduction} to find the independent generators of the group, $\mathcal{G}_{j}^{'}$. 
We use the implementation in Stim~\cite{Gidney_2021} for this step, which given the abelian group, $\mathcal{G}_{j}^{'}$, ignores the redundant generators and can output a stabilizer group, $\mathcal{S}_{j}$, by concatenating the independent generators with arbitrary independent generators.

\subsection{Graphs}\label{subsec:Gr}
In this section, we provide details of how to construct the corresponding graph state given the tableau representation of a stabilizer group.

A graph on $m$-vertices is defined by an adjacency $m\cross m$ binary matrix, $A$, where $A_{i,j}\ne0$ if there exists an edge between the $i$-th and $j$-th vertices.
Given a graph, one can then define a graph state as,
\begin{equation}
    \ket{\Psi_G} = \prod_{i<j} \text{CZ}_{i,j}^{A_{i,j}} \text{H}^{\bigotimes m} \ket{0}^{\bigotimes m},
\end{equation}
where CZ is the controlled-$\hat{\sigma}_{z}$ gate and H is the Hadamard gate.

It is well known that a stabilizer state, $\ket{\Psi_{s}}$, is local-Clifford equivalent to a graph state~\cite{Van_den_Nest_2004,zeng2007local}, $\ket{\Psi_{g}}$, that is there exists single-qubit unitaries, $\{\mathcal{V}_i\}$, such that
\begin{equation}
    \ket{\Psi_g} = \prod_i \mathcal{V}_i \ket{\Psi_s}.
\end{equation}
We follow the Algorithm 1 in Ref.~\cite{vijayan2022compilation} to find these local Clifford unitaries, $\{\mathcal{V}_i\}$, and the graph state $\ket{\Psi_g}$ using the tableau representation of a stabilizer group, $\mathcal{S}_i$.
One can then construct Clifford circuits, $\mathcal{U}_{z}$,
\begin{equation}
    \mathcal{U}_z = \prod_i \mathcal{V}_i \prod_{i<j} \text{CZ}_{i,j}^{A_{i,j}} \text{H}^{\bigotimes n}
\end{equation}
by using the graph state, $\ket{\Psi_g}$, and the local unitaries, $\{\mathcal{V}_i\}$, that simultaneously diagonalize all the operators, $\hat{P}$, in the corresponding group $\mathcal{G}_i$ as
\begin{equation}
    \hat{P}^{'} = \mathcal{U}_z^\dagger \hat{P} \mathcal{U}_z,
\end{equation}
where $\hat{P}^{'} \in \{ \hat{I}, \hat{\sigma}_{z}\}^{\bigotimes n}$ is a diagonal Pauli-string.
A procedure for constructing such circuits was also proposed in Ref.~\cite{miller2022hardware}, where the authors use the connectivity graph of the quantum device to construct hardware tailored diagonalization circuits by reformulating the problem as integer quadratically constrained program and using efficient algebraic solvers.

\begin{figure*}[htbp!]
    \centering
    \includegraphics[width=0.99\textwidth]{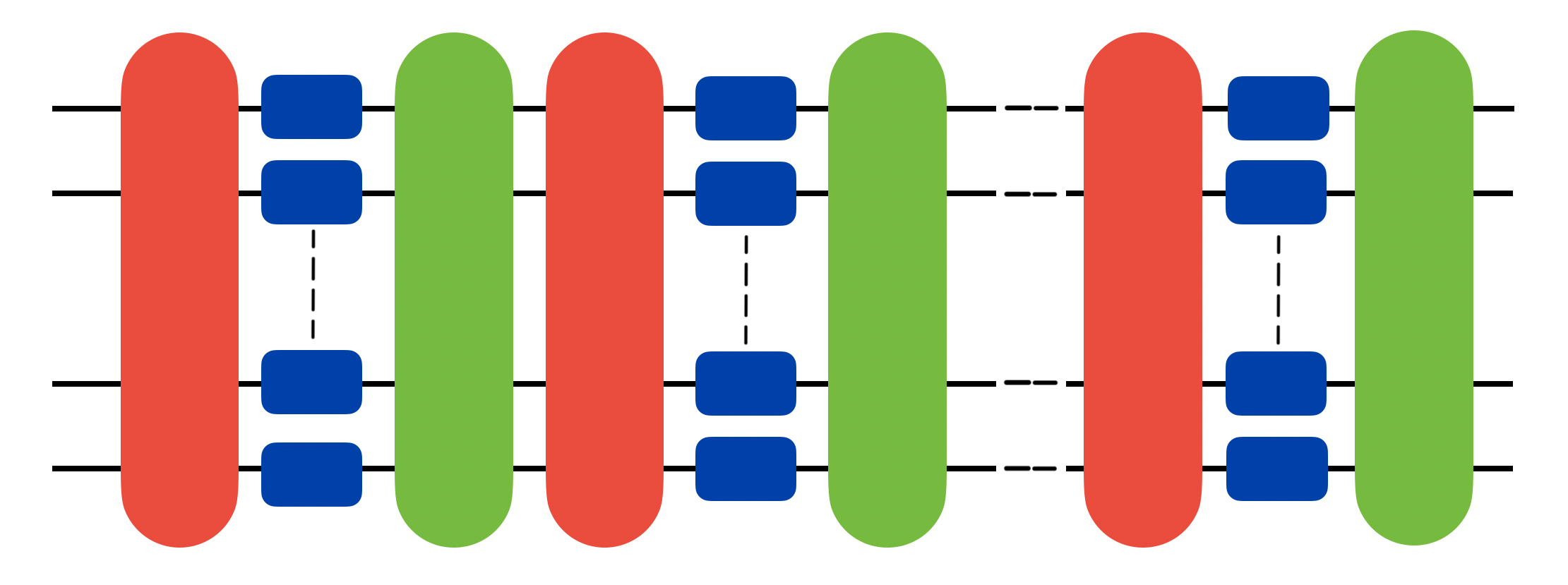} 
    \caption{\label{fig:HHEA} A schematic of a single layer of the Hamiltonian-based graph states ansatz (H-GSA). The red and green box represent Clifford circuits, $\mathcal{U}_{z_j}^{\dagger}$ and $\mathcal{U}_{z_j}$, respectively. An illustration of the Clifford circuit, $\mathcal{U}_{z_j}^{\dagger}$, is shown in Fig.~\ref{fig:GSC}. The blue boxes represents general single qubit rotation gate of the kind $ R_z(\alpha) R_y(\beta) R_z(\gamma_{j}) R_y(-\beta) R_z(-\alpha) $.} 
\end{figure*}

\subsection{Ans\"atze}\label{sec:ansatz}
In this section, we present the details of how to use the circuits $\mathcal{U}_z$ to construct an ansatz for approximating low-lying eigenvalues. 

An ansatz is a trial wavefunction, $\ket{\Psi(\theta)}$ defined as the state after application of a parameterized quantum circuit, $U(\theta)$, as
\begin{equation}
    \ket{\Psi(\theta)} = U(\theta) \ket{\Psi_0},
\end{equation}
where $\ket{\Psi_0}$ is some reference function.
They are used in variational algorithms to find the optimal value of an objective function.
In this article, we are interested in finding low-lying eigenvalues of a molecular Hamiltonian, $\hat{H}$, so our objective function is
\begin{equation}\label{eq:obj}
    E(\theta) = \bra{\Psi_0} U^\dagger(\theta) \hat H U(\theta) \ket{\Psi_0} = \bra{\Psi(\theta)} \hat H \ket{\Psi(\theta)},
\end{equation}
where $\theta$ are parameters that are optimized.

We can construct an ansatz using the molecular Hamiltonian, $\hat{H}$, as follows:
\begin{enumerate}
    \item Given $\hat{H}$, cluster the terms into commuting groups, $\{ \mathcal{G}_j \}$, by following the procedure in sec.~\ref{subsec:CG}.
    \item Order the groups as per the 1-norm of the terms in the group.
    \item Use the commuting groups, $\{ \mathcal{G}_j \}$, to construct stabilizer groups, $\{ \mathcal{S}_j \}$, by following the procedure in sec.~\ref{subsec:SG}.
    \item  Construct diagonalization circuits, $\mathcal{U}_{z_{j}}$, for all the stabilizer group, $\{ \mathcal{S}_j \}$, by following the procedure in sec.~\ref{subsec:Gr}.
    \item Combine the circuits together as per the order in step 2 to form a Hamiltonian-based graph states ansatz (H-GSA) of the form, 
    \begin{equation}
        \ket{\Psi(\theta)} = \prod_j \mathcal{U}_{z_j} R(\theta_j)  \mathcal{U}^{\dagger}_{z_j}  \ket{\Psi_0},
    \end{equation}
    where $R(\theta_j) = \bigotimes_{j}^{n} R_{\mathbf{v_j}}(\gamma_{j})$ is a layer of arbitrary single-qubit rotation gates and $\{\mathbf{v_j}, \gamma_{j}\} \in \{\theta_{i}\}$ are free parameters to be optimized.
    A schematic of the ansatz is shown in Fig.~\ref{fig:HHEA}.
\end{enumerate}

\begin{figure}[htbp!]
    \centering
    \includegraphics[width=0.5\textwidth]{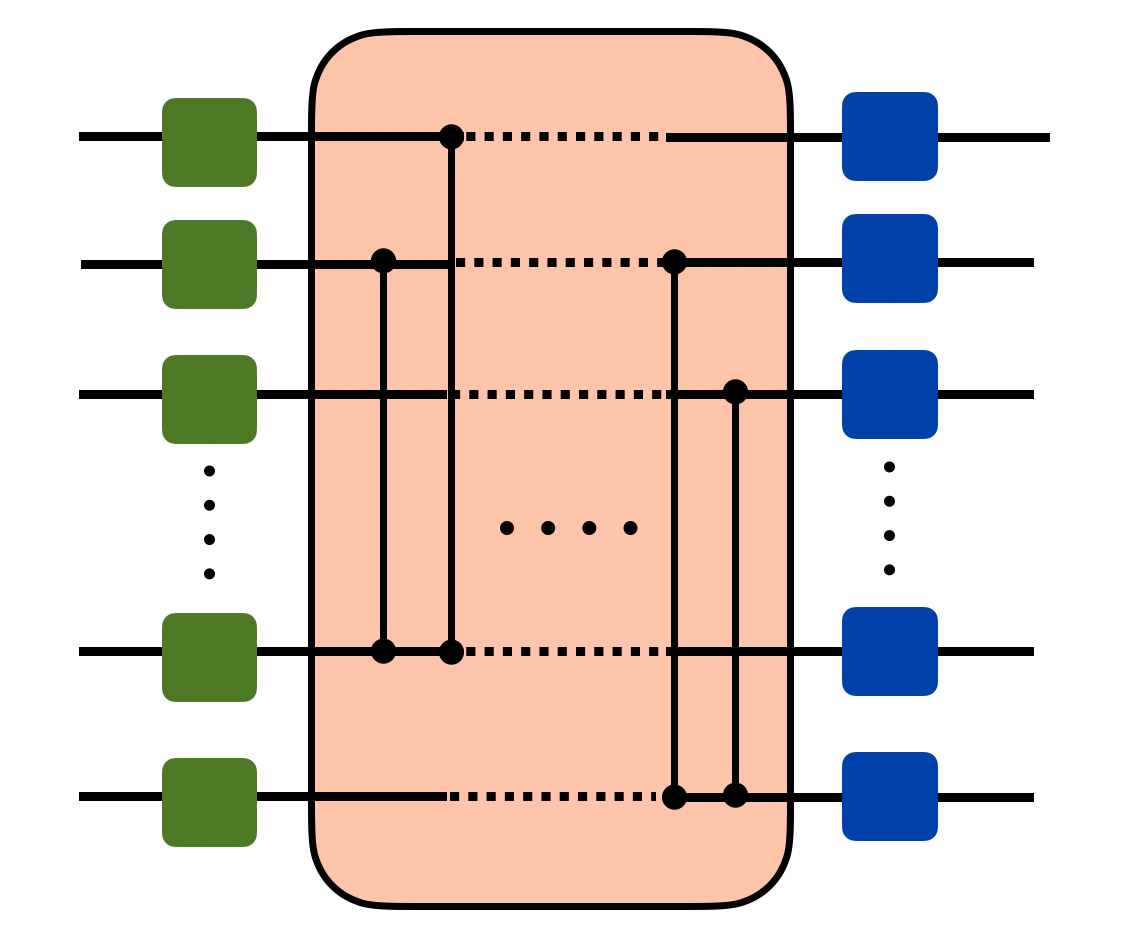} 
    \caption{\label{fig:GSC} A schematic of the Clifford circuits used in the Hamiltonian-based graph states ansatz (H-GSA). The pink box represent the circuit represent the layer of CZ-gates ($\prod_{i<j} \text{CZ}_{i,j}^{A_{i,j}}$) as per the adjacency matrix, $A$, of the graph state. The blue boxes represent Hadamard gates and the green boxes represent local Clifford gates, $\mathcal{V}_i$.} 
\end{figure}

A single rotation gate, $R_{\mathbf{v_j}}(\gamma_{j})$, in the layer, $R(\theta_{i})$, is a rotation by an angle, $\gamma_{j}$, about an arbitrary Bloch vector, $\mathbf{v_{j}} = \{ v_{x}, v_{y}, v_{z}\}$.
It is defined as
\begin{equation}\label{eq:rot_layer}
    R_{\mathbf{v_j}}(\gamma_{j}) = R_z(\alpha)R_y(\beta)R_z(\gamma_{j})R_y(-\beta)R_z(-\alpha),
\end{equation}
where $\alpha = \arctan(v_{y}/v_{x})$ and $\beta =\arccos(v_{z})$.

The number of free parameters in a single rotation gate is 3 and there are $n$ such gates in one rotation layer. 
There are $m$ layers corresponding to the $m$ commuting groups constructed above.
Thus, the total number of parameters in a single layer of the H-GSA circuit is $3mn$, which grows linearly in the number of qubits and the number of commuting groups.

We use the H-GSA defined above to minimize the objective function in Eq.~\ref{eq:obj} and find optimal parameters $\theta^{*}$, that approximates the ground state energy $E_{g}$
\begin{equation}
    \bra{\Psi(\theta^{*})} \hat H \ket{\Psi(\theta^{*})} \approx E_g.
\end{equation}

\section{Numerical Experiments}\label{sec:simulations}
In this section, we present results from different numerical simulations to showcase the applicability of the proposed ansatz.
We implement the proposed ansatz using Tequila~\cite{kottmann2021tequila} and use Qulacs~\cite{suzuki2021qulacs} as the backend for the execution of all the numerical simulations.
In all the simulations, we use the BFGS implementation of SciPy~\cite{virtanen2020scipy} for gradient based optimization of the objective function, $O(\theta)$.
The maximum number of iterations for all the numerical simulations is fixed to 200, unless stated otherwise.

\subsection{Trainability of H-GSA}\label{sec:train}
We first discuss the trainability aspect of the proposed ansatz.
It was shown in Ref.~\cite{grant2019initialization}, that the barren plateau phenomena can be resolved in quantum algorithms which uses deep parameterized quantum circuits, by choosing an initialization strategy where the circuit is a product of shallow identity blocks. 
The proposed ansatz by construction has such a decomposition, 
\begin{equation}
    U(\theta) = \prod_j U_i(\theta_j) = \prod_j \mathcal{U}_{z_j} R(\theta_j)  \mathcal{U}^{\dagger}_{z_j},
\end{equation}
where every block $U_i(\theta_j)$ is a conjugation of the non-Clifford circuit rotation layer, $R(\theta_j)$ by a Clifford unitary $\mathcal{U}_{z_j}$.
If we fix $\gamma_{j}$ in Eq.~\ref{eq:rot_layer} to be a very small value ($\sim$ 10$^{-6}$) then $U_j(\theta_j) \approx \hat{I}$, for arbitrary values of $\alpha$ and $\beta$. 
Thus, the proposed ansatz has the structure amenable for optimization with gradient-based methods as shown in Ref.~\cite{grant2019initialization}.

\subsection{Simulations for eigenvalue estimation}
We use the Bravyi-Kitaev transformation~\cite{bravyi2002fermionic} to convert the fermionic Hamiltonian (Eq.~\ref{eq:sec_q_ham}) of different molecules to the qubit Hamiltonian as in Eq.~\ref{eq:qubit_ham}.
We use the Hartree-Fock (HF) state as the initial state in all the numerical simulation carried out in this paper, however, one can use a better initial state as proposed in Ref.~\cite{anand2023leveraging}.
In all the simulation, we compare the result from our simulations to the exact energies calculated using the full configuration interaction (FCI) method.

We consider molecules of varying sizes for our numerical simulations: smaller molecules (4-6 qubits) - H$_2$ and LiH, medium-sized molecules (8-10 qubits) - H$_4$ and H$_2$O, and a larger molecule (12 qubits) - N$_2$. 
The results from these simulations are detailed in the following subsections.
All energy values are in Hartree (Ha) units and all bond length values are in Angstrom (\AA) units,  unless specified otherwise.

In all simulations, the variables $\boldsymbol{\alpha} = [\alpha_1,...,\alpha_N]$ and $\boldsymbol{\beta} = [\beta_1,..,\beta_N]$ are randomly initialized, while the value of $\boldsymbol{\gamma} = [\gamma_1,..,\gamma_N]$ is fixed at  1e-6, with $N=3mn$ representing the total number of variables in the circuit.
We conduct multiple experiments with different initial values and demonstrate that the method consistently converges to a solution, regardless of the starting points, thereby confirming the trainability of the proposed approach.

\subsubsection{H$_2$ and LiH molecules}\label{sec:h2andLih}
We use the full minimal basis set (STO-3G) for the hydrogen molecule and use only an active space of the lithium hydride molecule.
The hydrogen molecule in this representation has 2-electrons in 4 spin orbitals and the lithium hydride molecule has 2-electrons in 6 spin orbitals.
We then construct the objective function for these molecules by using the H-GSA as described in sec.~\ref{sec:ansatz}.

\begin{figure}[htbp!]
    \centering
    \begin{tabular}{c c}
    \toprule
    a) H$_{2}$ molecule   & b)  LiH molecule \\
    \midrule
    \multicolumn{2}{c}{\textbf{Energy}}\\
    \midrule
    \includegraphics[width=0.5\columnwidth]{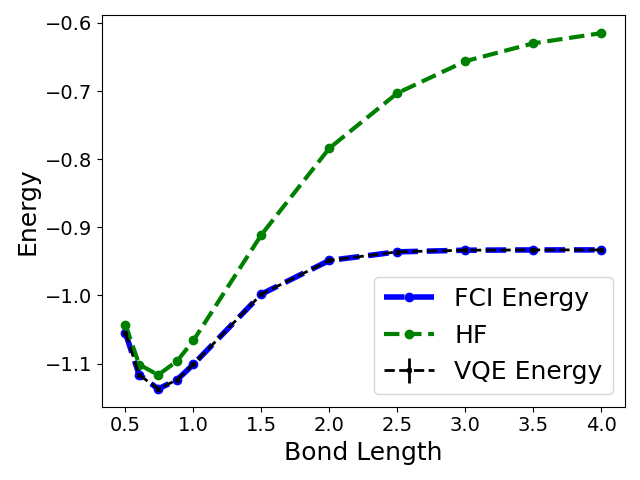} & 
    \includegraphics[width=0.5\columnwidth]{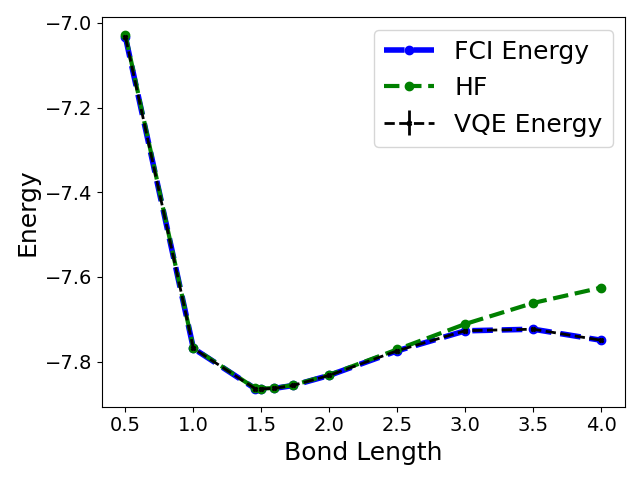}\\
    \midrule
    \multicolumn{2}{c}{\textbf{Error in energy}}\\
    \midrule
    \includegraphics[width=0.5\columnwidth]{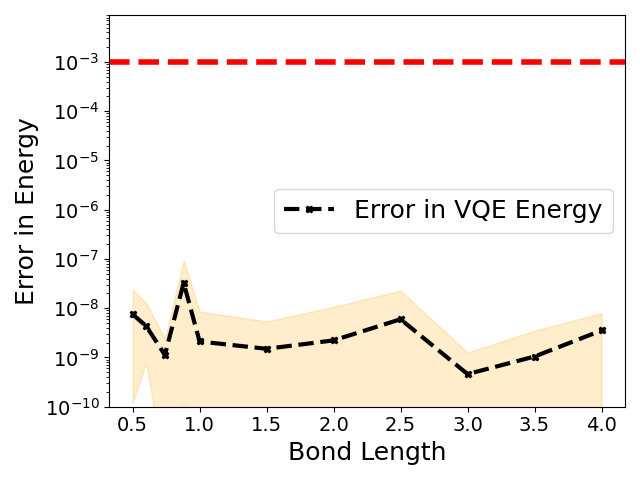} & 
    \includegraphics[width=0.5\columnwidth]{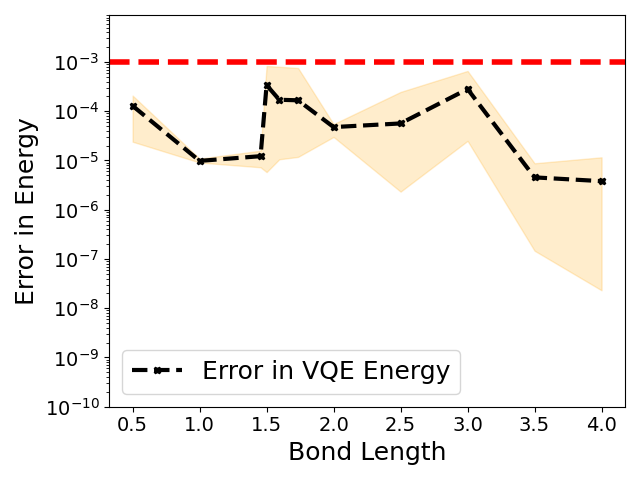}\\
    \midrule
    \end{tabular}
    \caption{\label{fig:H2andLiH} VQE results from optimization of ground state energies for H$_{2}$ and LiH molecule with different geometries using a single layer of the H-GSA circuit. 
    Top panel: The average value of the energies from the different simulation runs are plotted.
    Bottom panel: The average error value in plotted using the solid blue line and the shaded region represent the area between the smallest and the largest error value.
    The red line denotes an error value of 1e-3.} 
\end{figure}

We randomly sample five different parameter values while maintaining the identity block initialization scheme discussed in sec.~\ref{sec:train} and run different numerical simulations to estimate ground state energies of these molecules.
The results from these numerical simulations are presented in Fig.~\ref{fig:H2andLiH}.

It can be seen from the plots in Fig.~\ref{fig:H2andLiH} that using a single layer of the H-GSA circuit, for all the numerical simulations with different random parameter initializations, we are able to get well within the desired accuracy value (1e-3) of the exact ground state (FCI) energies of the both the molecules.
This suggest that the proposed ansatz is expressive enough to estimate the ground state energies of small molecules, while being trainable.

An example demonstration of the proposed ansatz for calculating eigenvalue of H$_2$ molecule is shown here~\cite{github_Ham_MVHA}.

\subsubsection{H$_4$ and H$_2$O molecules}
\begin{figure}[htbp!]
    \centering
    \begin{tabular}{c c}
    \toprule
    a) H$_{4}$ molecule   & b)  H$_{2}$O molecule \\
    \midrule
    \multicolumn{2}{c}{\textbf{Energy}}\\
    \midrule
    \includegraphics[width=0.5\columnwidth]{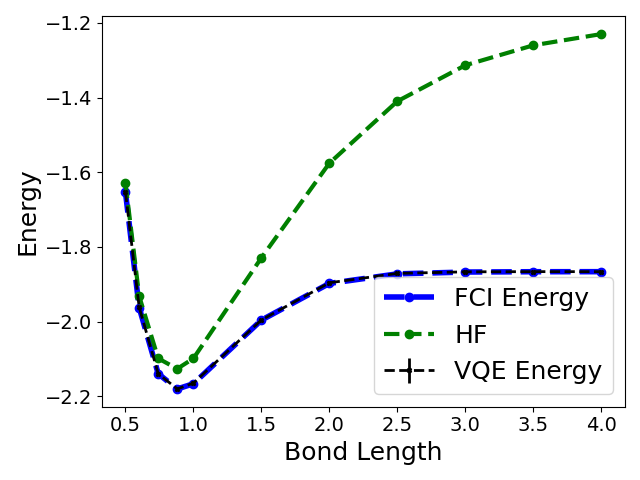} & 
    \includegraphics[width=0.5\columnwidth]{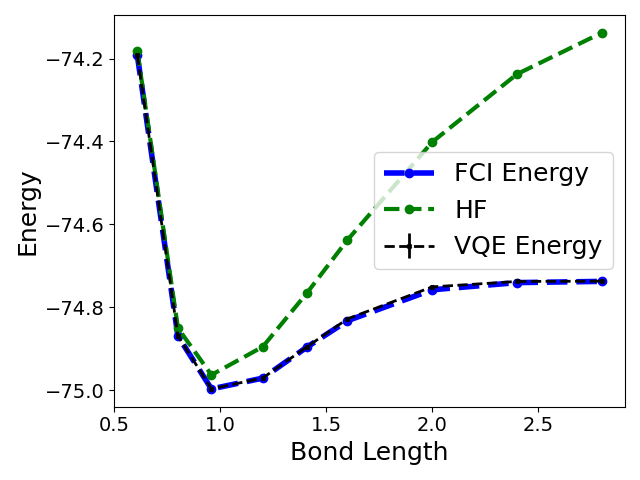}\\
    \midrule
    \multicolumn{2}{c}{\textbf{Error in energy}}\\
    \midrule
    \includegraphics[width=0.5\columnwidth]{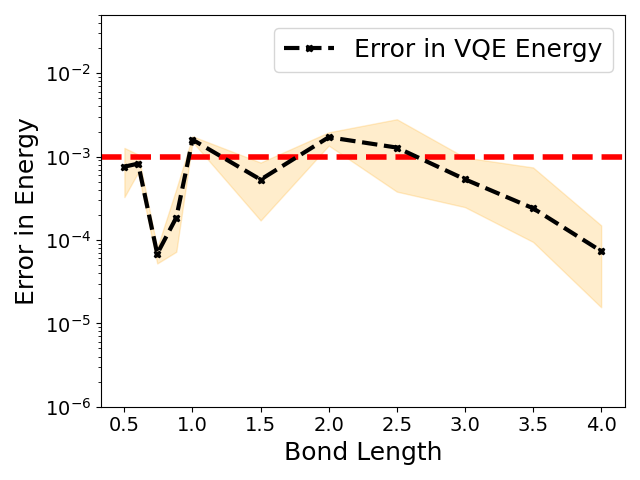} & 
    \includegraphics[width=0.5\columnwidth]{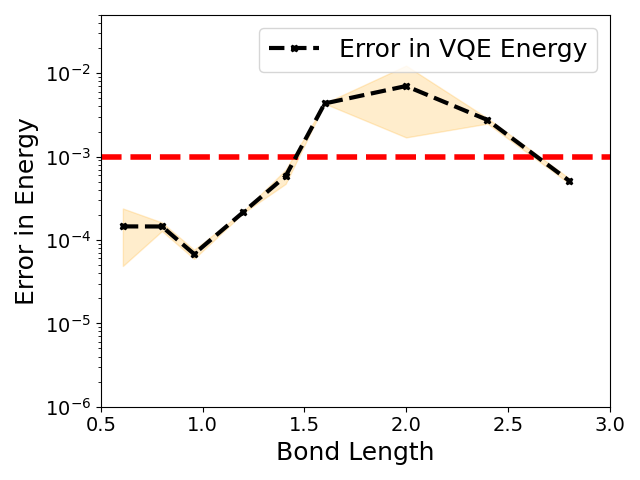}\\
    \midrule
    \end{tabular}
    \caption{\label{fig:H4andH2O} VQE results from optimization of ground state energies for H$_{4}$ and H$_{2}$O molecule with different geometries using a single layer of the H-GSA circuit.
    Top panel: The average value of the energies from the different simulation runs are plotted.
    Bottom panel: The average error value in plotted using the black line and the shaded region represent the area between the smallest and the largest error value.
    The red line denotes an error value of 1e-3.} 
\end{figure}

We use the full minimal basis set (STO-3G) for the linear hydrogen chain H$_4$ and an active space of the minimal basis set for the water molecule.
The hydrogen chain has 4-electrons in 8 spin orbitals and the water molecule has 6-electrons in 10 spin orbitals.
We use the proposed H-GSA for estimating the ground state energies of these molecules and run similar numerical simulations as in sec.~\ref{sec:h2andLih}.
The results from the numerical simulations are plotted in Fig.~\ref{fig:H4andH2O}.

From the plots in Fig.~\ref{fig:H4andH2O}, it can be observed that the energies from the numerical simulations for most molecular configurations are within the desired chemical accuracy (1e-3 Ha) from the exact FCI energy.

For the $H_{4}$ molecule, there are a few points on the full PES where the smallest error exceeds 1e-3 Ha. 
We attribute this to the limited number of iterations (200) in the simulations, during which some runs failed to converge to the true ground state and instead converged to local minima.

For the case of water molecule, it can be seen from the plots in Fig.~\ref{fig:H4andH2O} that the error in energies for molecules with configuration near equilibrium geometry are within the desired chemical accuracy (1e-3 Ha). 
The errors keep getting larger than the desired accuracy value of 1e-3 for stretched configurations before getting close to the desired accuracy for bond length close to dissociation.
This is an expected behavior where a layer of the proposed ansatz is not expressive enough for geometries of water molecule where the ground states have high dynamic correlations.

\begin{figure}[htbp!]
    \centering
    \begin{tabular}{c c}
    \toprule
    a) H$_{4}$ molecule   & b)  H$_{2}$O molecule \\
    \midrule
    \multicolumn{2}{c}{\textbf{Error in energy}}\\
    \midrule
    \includegraphics[width=0.5\columnwidth]{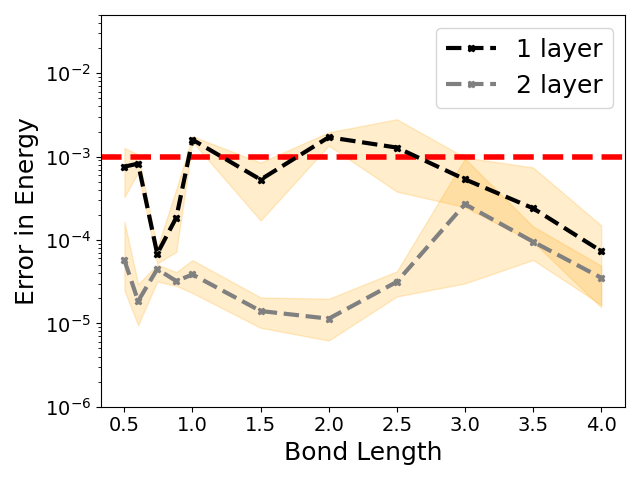} & 
    \includegraphics[width=0.5\columnwidth]{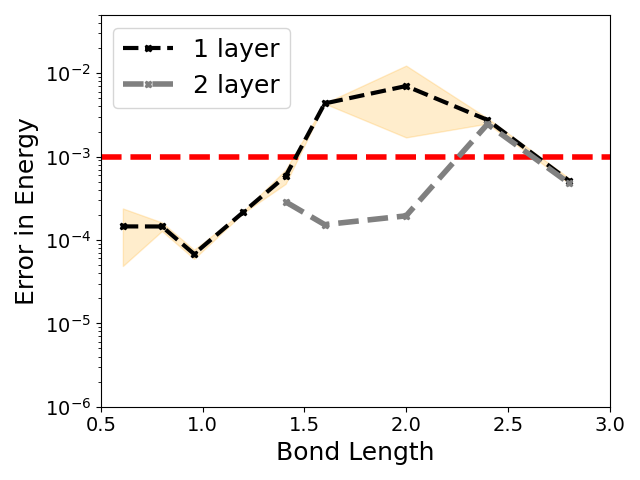}\\
    \midrule
    \end{tabular}
    \caption{\label{fig:12_layr_comp} VQE results from optimization of ground state energies for H$_{4}$ and H$_{4}$ molecule with different geometries using one and two layers of the H-GSA circuit. 
    The average error value with 1 and 2 layers in plotted using the black and gray line, respectively.
    The shaded region represent the area between the smallest and the largest error value.
    The red line denotes an error value of 1e-3.} 
\end{figure}

We believe that repeating the ansatz multiple times can lead to improved convergence, even for these regimes, as observed in simulations from other studies~\cite{anand2023leveraging, choquette2020quantum, anand2022exploring} employing Hamiltonian-based ansätze.
To test this hypothesis, we conducted additional numerical simulations using two layers of the H-GSA circuit. 
The results are presented in Fig.~\ref{fig:12_layr_comp}.
From these plots, it is evident that the final energies obtained from these simulations are within the desired chemical accuracy (1e-3 Ha) of the true ground state energy, except for one configuration of the water molecule. 
These results support the hypothesis that increasing the number of layers can improve convergence.
We further believe that these results could be enhanced by either increasing the number of layers in the ansatz or employing advanced optimization techniques, such as those proposed in Ref.~\cite{Cervera_Lierta_2021}.

\begin{figure*}
    \centering
    \begin{tabular}{c c c c}
    \toprule
    a) 0.5 \textup{\AA}   & b)  1.09 \textup{\AA} &c) 2.0 \textup{\AA}   & d)  3.0\textup{\AA}\\
    \midrule
    \multicolumn{4}{c}{\textbf{Energy}}\\
    \midrule
    \includegraphics[width=0.5\columnwidth]{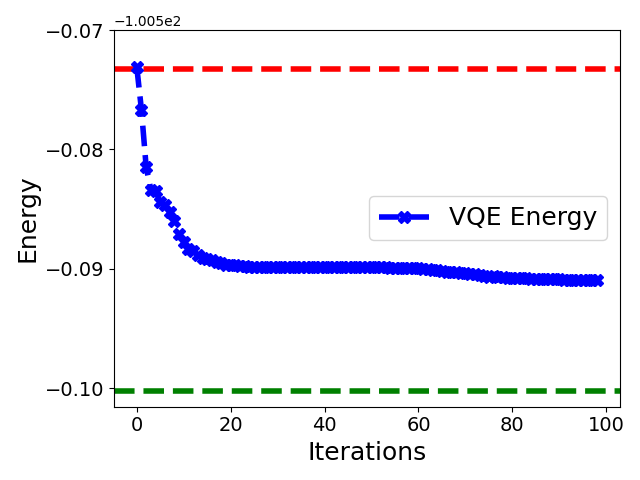} & 
    \includegraphics[width=0.5\columnwidth]{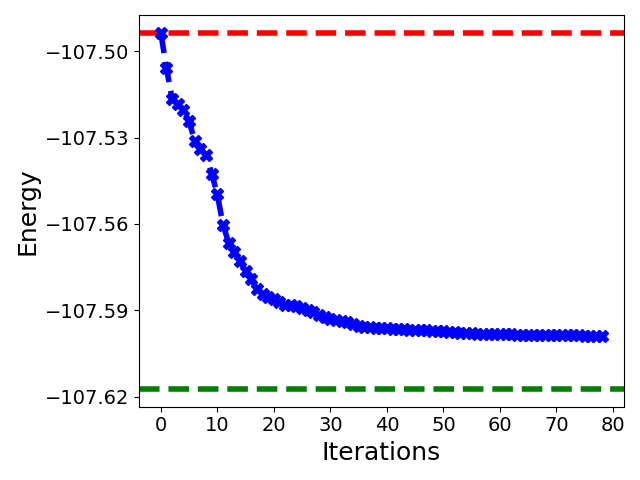} &
    \includegraphics[width=0.5\columnwidth]{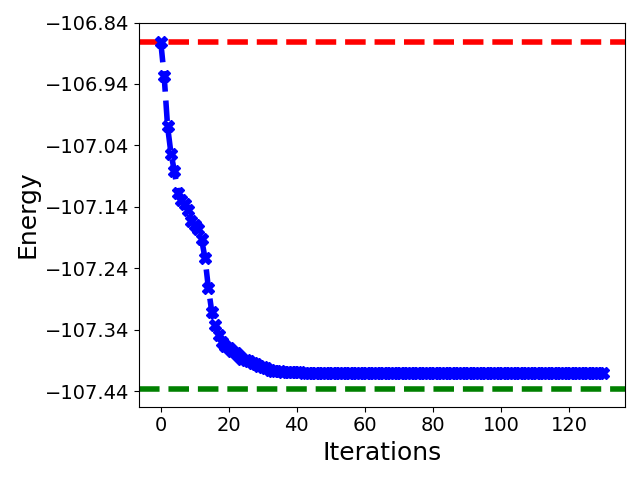} & 
    \includegraphics[width=0.5\columnwidth]{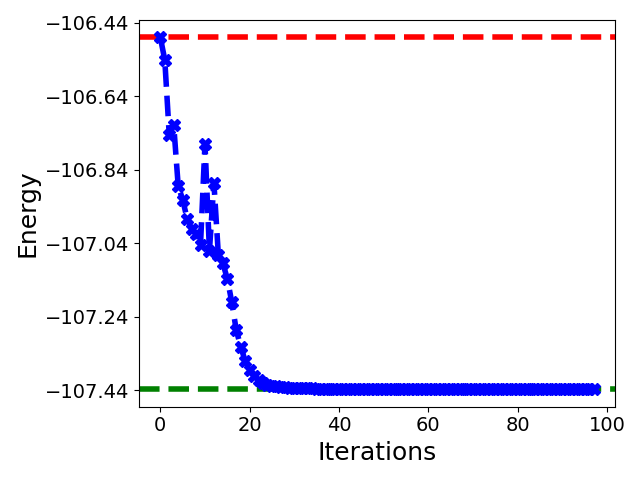}\\
    \midrule
    \multicolumn{4}{c}{\textbf{Error in energy}}\\
    \midrule
    \includegraphics[width=0.5\columnwidth]{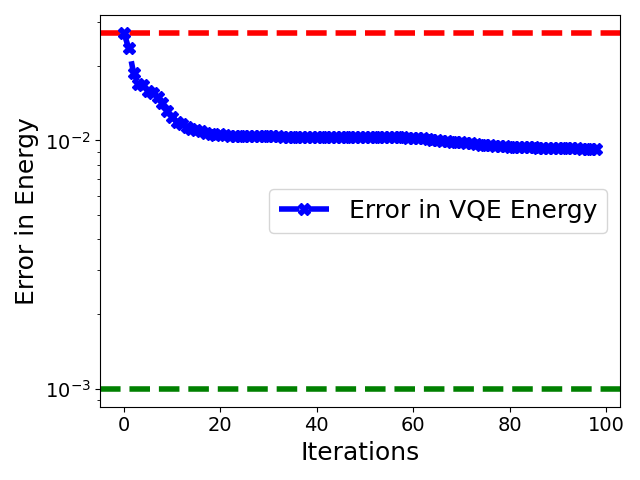} & 
    \includegraphics[width=0.5\columnwidth]{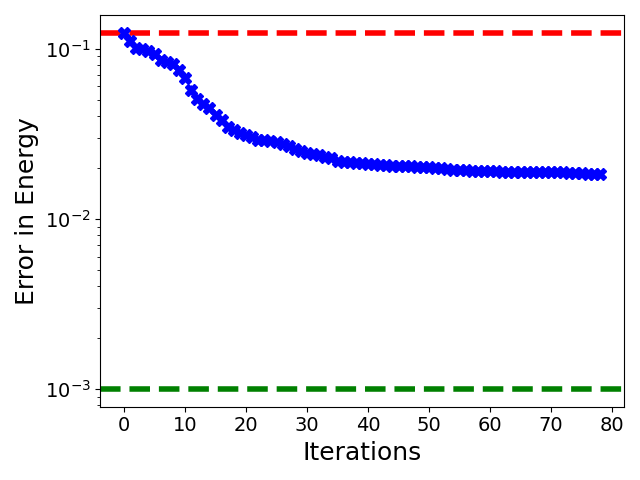} &
    \includegraphics[width=0.5\columnwidth]{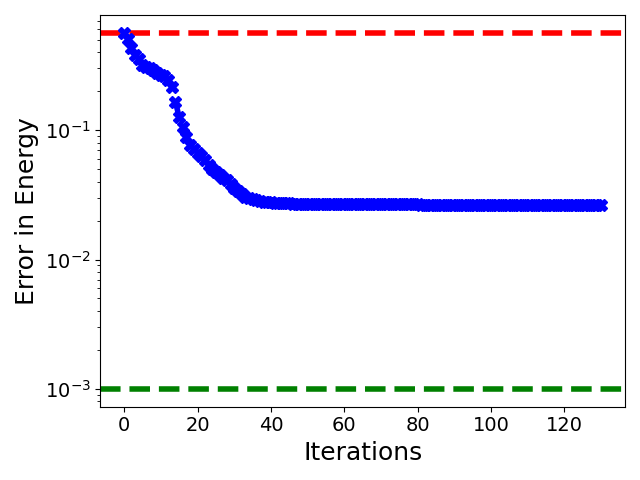} & 
    \includegraphics[width=0.5\columnwidth]{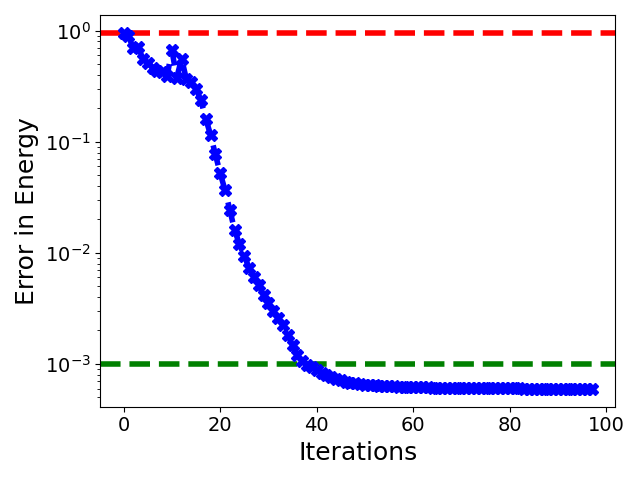}\\
    \bottomrule
    \end{tabular}
    \caption{\label{fig:N2} Optimization trajectories from the numerical simulation for calculating ground state energies for N$_{2}$ molecule using different geometries with a single layer of the H-GSA circuit.
    Top panel: The red and green line correspond to the Hartree-Fock energy and the exact ground state (FCI) energy. The blue line correspond to the optimization trajectory using a single layer of H-GSA circuit.    
    Bottom panel: The red line correspond to the error corresponding to the HF state. The green line correspond to an error value of 1e-3. The blue line correspond to error in energy during the optimization runs.} 
\end{figure*}

\subsubsection{N$_2$ molecule}
In this section, we use an activate space of the minimal basis set (STO-3G) for the nitrogen (N$_2$) molecule, where the molecule has 6-electrons in 12 spin orbitals. 
We use the proposed H-GSA circuit to calculate eigenvalues of the molecule at four different bond lengths, 0.5, 1.09. 2.0, and 3.0 $\textup{\AA}$.
We plot the optimization trajectory for these simulations in Fig.~\ref{fig:N2}.

It can be observed from the plots in Fig.~\ref{fig:N2} that we are able to carry out optimization for all configurations. 
However, we only converge to the exact ground state energy within desired accuracy for the configuration where the nitrogen molecule is close to dissociation.
For all the other configurations we observe that the energy converges to a value larger than the true ground state energy.
This is expected behavior as these regimes have ground states with higher dynamic correlation compared to the one close to dissociation.
We believe that a layer of the H-GSA circuit is not expressive enough to represent these highly correlated ground state wavefunctions and we anticipate better energy estimates with repeated layer of the ansatz.

\subsection{Comparison to other ans\"atze}
In this section we analyze and compare the gate count complexity and the number of parameters in the proposed ansatz to previously proposed Hamiltonian based schemes.
Given a Hamiltonian of the form in eq.~\ref{eq:qubit_ham}, an initial state, $\ket{\Psi_0}$ and a set of parameters, $\{\theta\}$, we briefly mention the previous schemes:
\begin{enumerate}
    \item Traditional variational Hamiltonian ansatz (T-VHA) -
    The ansatz proposed in Ref.~\cite{wecker2015progress} has the form:
    \begin{equation}
        \ket{\Psi(\theta)} = \prod_{k=1} e^{-i\theta_{k}\hat{P}_{k}}   \ket{\Psi_0}.
    \end{equation} 
    \item Simultaneously diagonalizable variational Hamiltonian ansatz (D-VHA) - This ansatz has the form:
    \begin{equation}\label{eq:VHA_}
    \ket{\Psi(\theta)} = \prod_{k=1}^{m} \mathcal{U}^{\dagger}_k (\prod_{l=1}^{m_k} e^{-i\theta_{kl} \mathcal{U}_k \hat{P}_{kl} \mathcal{U}^{\dagger}_k}) \mathcal{U}_k \ket{\Psi_0},
    \end{equation}    
    where, $\mathcal{U}_k$ are Clifford circuits that diagonalize the Pauli-strings $\{\hat{P}_{kl}\}$.
    \item Modified variational Hamiltonian ansatz (M-VHA) - This ansatz proposed in Ref.~\cite{anand2023leveraging} has the form:
    \begin{align}
    \nonumber
    \ket{\Psi(\theta)}
     &= \prod_{i}^{m} \mathcal{U}_i^{\dagger} \bigotimes_{j=1}^{n} \textbf{R}_{x_{j}}(\theta_{x_{i,j}})\textbf{R}_{y_{j}}(\theta_{y_{i,j}}) \\
     & \textbf{R}_{z_{j}}(\theta_{z_{i,j}})  \mathcal{U}_i \ket{\Psi_0},
    \end{align}
    where, $\textbf{R$_x$}, \textbf{R$_y$}, \textbf{R$_z$}$ are single qubit rotations and $\mathcal{U}_k$ are Clifford circuits.
\end{enumerate}

It should be noted that the structure of the M-VHA and H-GSA circuits are very similar, with the main difference between them being the structure of the Clifford circuits used to diagonalize the individual groups.

We now look at the two qubit gate count for all the ans\"atze mentioned above for different molecules of size up to 20-qubits and compile them in Tab.~\ref{tab:VHAvsMVHA}.
It can be observed from the table that the gate count for the proposed H-GSA circuit is constantly an order magnitude smaller than all the previously proposed Hamiltonian-based schemes.
It can be seen from Refs.~\cite{anastasiou2024tetris, Kottmann_2022} the gate count is also comparable to that of other ans\"atze proposed specifically for molecular Hamiltonians.
This indicates that the ansatz proposed here is very compact.

\begin{table}[htbp!]
    \centering
    \begin{tabular}{c| c | c | c | c}
        \hline
        \multicolumn{1}{c}{Molecule (N$_e$, N$_q$)} & \multicolumn{4}{c}{Number of 2-qubit gates} \\
        \hline
        \hline
        & T-VHA & D-VHA & M-VHA & H-GSA \\
        \hline
        H$_2$ (2,4) & 44 & 56 & 16 & 4 \\ 
        \hline
        LiH (2,6) & 262 & 356 & 144 & 18 \\ 
        \hline
        H$_4$ (4,8) & 1320 & 1636 & 722 & 90 \\ 
        \hline
        H$_2$O (6,10) & 1992 & 2582 & 1128 & 170 \\ 
        \hline
        N$_2$ (6,12) & 2168 & 2605 & 1056 & 168 \\ 
        \hline
        BeH$_2$ (6,14) & 6276 & 8862 & 4277 & 690 \\ 
        \hline
        BH$_3$ (8,16) & 17125 & 24935 & 13245 & 1995 \\ 
        \hline
        H$_{10}$ (10,20) & 104130 & 136378 & 67696 & 9235 \\ 
        \hline
        \hline
    \end{tabular}
    \caption{A table containing the two-qubit gate complexity of a single layer of different Hamiltonian based ans\"atze. The number (N$_e$, N$_q$) in parenthesis correspond to the number of electrons and qubits, respectively.}
    \label{tab:VHAvsMVHA}
\end{table}

Next, we compute the number of parameters in the proposed ansatz and compare it to other previously proposed Hamiltonian based circuits.
We compute the values for all the different ans\"atze and compile them in Tab.~\ref{tab:VHAvsMVHA_params}.
It can be seen that the number of parameters nearly doubles as compared to the previous schemes.
However, we point out that the number only grows linearly (see Sec.~\ref{sec:ansatz}) both, with the size of the system and the number of commuting groups in the Hamiltonian.

\begin{table}[htbp!]
    \centering
    \begin{tabular}{c| c | c}
        \hline
        \multicolumn{1}{c}{Molecule (N$_e$, N$_q$)} & \multicolumn{2}{c}{Number of Parameters} \\
        \hline
        \hline
        & T-VHA/D-VHA & M-VHA/H-GSA \\
        \hline
        H$_2$ (2,4) & 14  & 12 \\ 
        \hline
        LiH (2,6) & 61  & 108 \\ 
        \hline
        H$_4$ (4,8) & 184  & 198 \\ 
        \hline
        H$_2$O (6,10) & 251  & 476 \\ 
        \hline
        N$_2$ (6,12) & 246 &  364 \\ 
        \hline
        BeH$_2$ (6,14) & 665  & 1512 \\ 
        \hline
        BH$_3$ (8,16) & 1520 & 3174 \\ 
        \hline
        H$_{10}$ (10,20) & 7148 & 8200 \\ 
        \hline
        \hline
    \end{tabular}
    \caption{A table containing the number of parameters in a single layer of different Hamiltonian-based ans\"atze. The number (N$_e$, N$_q$) in parenthesis correspond to the number of electrons and qubits, respectively.}
    \label{tab:VHAvsMVHA_params}
\end{table}

\section{Conclusion}\label{sec:conclusion}
We have presented a new method for constructing Hamiltonian-based quantum circuits for estimating eigenvalues of different Hamiltonians.
The quantum circuit is constructed using graph-based diagonalization circuits and layers of arbitrary single qubit rotation gates.
We note that the structure of the circuit allows for it to be initialized as a product of shallow identity blocks.
The initialization strategy leads to the proposed ansatz having good trainability properties while being very shallow but with increased number of free parameters.

We then validate this using numerical simulations and find that with only a single layer of the ansatz we are able to get a very good estimate of the ground state energy for different configurations of various molecules.
However, we observe that as we simulate larger systems a single layer of the ansatz is not expressible enough, but we expect repeating the ansatz can lead to better energy estimates.
The proposed method serves as a significant step towards constructing shallow circuits for arbitrary Hamiltonians with applications in chemistry and physics.

Although we have very shallow circuits, we need to carry out further studies to estimate resources for practical executions of these circuits on existing hardware by considering the effect of sampling error and finite precision of rotation angles on the number of circuit executions for a given accuracy of estimating eigenvalues. 
Other directions that need to be worked out is the dependence of optimization procedure on the performance of our methods. 
Further numerical work may help answer some of these questions.

\section*{Acknowledgements}
This work was supported by the National Science Foundation (NSF) Quantum Leap Challenge Institute of Robust Quantum Simulation (QLCI grant OMA-2120757).
A.A. also acknowledges support by the National Science Foundation (Grant No. DMS-1925919), as part of the work was done when he was visiting the Institute for Pure and Applied Mathematics (IPAM).

\bibliography{main.bib}

\end{document}